\documentstyle[11pt,newpasp,twoside,epsf]{article}
\newcommand{\be}{\begin{equation}}
\newcommand{\ee}{\end{equation}}
\newcommand{\bsf}[1]{\mbox{\begin{bfseries}\textsf{{#1}}\end{bfseries}}}
\newcommand{\nbsf}[1]{\mbox{\textsf{{#1}}}}
\def\simless{\mathbin{\lower 3pt\hbox
   {$\rlap{\raise 5pt\hbox{$\char'074$}}\mathchar"7218$}}}
\def\simgreat{\mathbin{\lower 3pt\hbox
   {$\rlap{\raise 5pt\hbox{$\char'076$}}\mathchar"7218$}}}   

\def\edcomment#1{\iffalse\marginpar{\raggedright\sl#1\/}\else\relax\fi}
\reversemarginpar

\begin{document}
\title{Decomposition of the Visible and Dark Matter Mass Profiles in the
Einstein Ring 0047$-$2808}
\author{Simon Dye \& Steve Warren}
\affil{Astrophysics Group, Imperial College London, Blackett Labs, Prince
Consort Road, London, SW7 2BW, U.K.}

\begin{abstract}
Using our semi-linear inversion method, we measure the mass profile of
the lens galaxy in the Einstein ring system 0047$-$2808. The lens is
modelled as a baryonic component following the observed light, embedded
in a dark matter generalised NFW halo. The semi-linear method makes
full use of the information content in the ring image. We determine an
unevolved B-band mass to light ratio for the baryons of
$3.05^{+0.45}_{-0.86} \,h_{65}\,M_{\odot}/L_{B\odot}$ (95\% CL),
accounting for 65\% of the total projected mass within the radius of
$1.16''$ traced by the ring. This result is obtained without need of
dynamical measurements. The inner logarithmic slope of the halo is
found to be $0.87^{+0.69}_{-0.61}$ (95\% CL). We find that the halo is
fairly well aligned with the light but has only half the ellipticity.
\end{abstract}

\section{Introduction}

Gravitational lensing provides a simple and well understood tool for
the determination of galaxy mass profiles. It has the advantage that
the deflection angle of a photon passing a massive object is
independent of the dynamical state of the deflecting mass. Dynamical
methods for measuring galaxy mass profiles rely on a number of
assumptions that can be difficult to test, such as dynamical
equilibrium, circular orbits of gaseous disks, and, frequently,
spherical symmetry of the mass distribution.

In contrast to weak lensing, where a statistical detection of
galaxy-galaxy lensing can place limits on mass profiles (eg. Hoekstra,
Yee, \& Gladders 2003), strong lensing involves the analysis of
individual systems and allows smaller scales to be probed. In
strong lens systems, multiple images of a background source are
observed.  Once the redshift of the source and lens galaxy are known,
mass models for the lens can be fit by comparing predicted image
positions with those measured. Strong lens systems where the source
has extended structure give rise to multiple arc-shaped images. These
arcs trace out the Einstein ring which, as Kochanek, Keeton, \& McLeod
(2001) show, allows stronger constraints to be placed on the lens mass
profile.  Unlike strong lens systems where the source is point-like,
the analysis of Einstein rings requires modelling of the source
surface brightness distribution. In this way, the properties of both
the source and the lens are minimised to give the best fit to the
observed ring.

In the analysis of Einstein rings, the way in which the source is
modelled has a significant effect. Assuming a simple analytical
surface brightness profile for the source can bias the lens model
solution. Real sources typically have complicated structure and an
over-simplified source forces the fitted lens model to compensate.  A
non-parametric form, such as one in which the source surface
brightness is pixelised, almost entirely removes this problem but
creates a new one: The surface brightness in each source pixel must be
adjusted in addition to the lens model parameters to find the minimum
thus hugely slowing the procedure.

Wallington, Kochanek, \& Narayan (1996) describe a method based on
maximum entropy that uses a pixelised source surface brightness
distribution. This has been applied by Wayth et al. (2003) to the
Einstein ring 0047$-$2808. At the heart of the method are two nested
minimisation cycles. For a given lens model stipulated by the outer
cycle, the surface brightnesses in the source pixels are adjusted in
the inner cycle until an image is obtained which gives the best fit to
the observed ring image. The inner cycle is a slow non-linear search
for the minimum requiring many iteration to reach convergence.

In a previous paper (Warren \& Dye 2003), we developed a new technique
called the semi-linear method. The semi-linear method uses a pixelised
source plane but replaces the inner cycle of the maximum entropy
method with a single-step linear inversion. This has the advantage that it
is very fast and, more importantly, that the correct source minimum is
guaranteed for a given lens model. The possibility of minimising to an
incorrect source, thereby biasing the minimised lens model, is thus
completely removed.

In the work presented here, we apply the semi-linear method to WFPC2
observations of the Einstein ring 0047$-$2808. We model the lens as a
baryonic component embedded in a dark halo and show how the
contribution from each component can be determined to allow
measurement of the baryonic M/L.

\section{The semi-linear reconstruction method}
\label{sec_method}

We refer the reader to Warren \& Dye (2003) for a complete description
of the semi-linear method. Here, we give an outline.

The inversion requires that both the source plane and the
image plane are pixelised. This pixelisation can be of a completely
general fashion in terms of the distribution of pixel sizes and
locations. Labelling the surface brightness in reconstructed source
pixel $i$ as $s_i$, the flux in observed image pixel $j$ as $d_j$
and the $1\sigma$ error on this flux as $\sigma_j$, then for a given
lens model, the vector of source surface brightnesses which best fits
the observed ring is given by
\be
\label{eq_reg_chisq_soln}
\bsf{s}=\left(\bsf{F}+\lambda\bsf{R}\right)^{-1}\,\bsf{t}
\ee
where
\be
\nbsf{F}_{kl}=\sum_{j=1}^N f_{kj}f_{lj}/\sigma_j^2\, , \quad
\nbsf{R}_{kl}=2\sum_{i=1}^M r_{ik}r_{il} \quad {\rm and} \quad
t_i=\sum_{j=1}^N f_{ij}d_{j}/\sigma_j^2.
\ee
The quantity $f_{ij}$ is the contribution to the flux in image pixel $j$
from source pixel $i$ according to the given lens model and allowing
for smearing by the observed image PSF. The solution is regularised by
the matrix $\bsf{R}$, the form of which is selected by the user. This
is chosen to reduce noise in the reconstructed source by penalising
spikes with the regularisation weight $\lambda$ (see Warren \& Dye 2003).

Equation (1) gives the best fit source for a fixed lens model.  This
is the linear part of the semi-linear method. The non-linear part
comes from the need to adjust the lens parameters in an outer
iterative cycle until the best global fit is obtained.  At each step
in the outer cycle, a new set of quantities $f_{ij}$ must be
calculated.

\section{Data}

Full details of the HST observations and data reduction steps are given
in Wayth et al. (2003).

The data consist of drizzled images of the Einstein ring 0047$-$2808
observed with the WFPC2 instrument. The source in this system is a
star-forming galaxy at redshift $z=3.595$. The galaxy has
strong Ly$\alpha$ emission (Warren et al. 1996), hence observations
were carried out in the F555W filter to maximise the ring:lens flux
ratio. The fully reduced, drizzled image ($2 \times 2$ dither
positions per full WFPC2 pixel) is shown in the left of Figure 1.

\begin{figure}[h]
\plotone{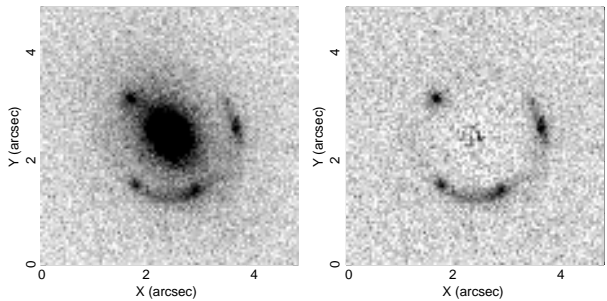}
\caption{\small Drizzled WFPC2 observation of 0047$-$2808 in F555W filter.
{\em Left:} Original image {\em Right:} Ring
with lensing galaxy subtracted (small central residuals not
significant)}
\end{figure}

Since the semi-linear method requires an image of the ring only, the
lens galaxy had to be first removed.  The best fitting surface
brightness distribution for the lens galaxy was found to be a Sersic
profile (Sersic 1968) with a central point source (Wayth et al.
2003). The plot on the right of Figure 1 shows the resulting Einstein
ring after the lensing galaxy has been subtracted. The residue at the
centre of the subtracted image is insignificant Poisson noise.

\section{Modelling procedure}

In our dual component lens model, we wish to constrain the inner
slope of the density profile of the dark halo and the baryonic mass-to-light
ratio.

The baryonic component of the lens model is assumed to follow the
light of the lens galaxy, fixed by the fitted Sersic + point source
profile.  The lens deflection angle due to the baryons is therefore
computed from this light profile scaled by the B-band baryonic mass
to light ratio $\Psi$ (in units of $h_{65}\,M_{\odot}/L_{B\odot}$). 
The surface density of the Sersic profile
in units of $M_{\odot}/\Box''$, is
\be
\label{eq_sersic_smd}
\Sigma_b(\eta) = \Psi L_{1/2} \exp\{-5.90[(\eta/1.31'')^{0.32}-1]\}
\ee
and the mass of the central point source,
\be
M_{p}=3.09\times 10^9 \,h_{65}^{-2}\,L_{\odot} \Psi,
\ee
expressed here in units of $M_{\odot}$. $\Psi$ is the only free
parameter describing the baryonic contribution to the total lens mass.
The half light luminosity of the Sersic profile is $L_{1/2}=1.99\times
10^9 \,h_{65}^{-2} L_{B\odot}/\Box''$ and the ellipse co-ordinate
$\eta^2=x^2+y^2/q^2$ has a measured axis ratio $q=b/a=0.69$. The
orientation of the major axis to the vertical in a counter-clockwise
direction is $35^{\circ}$.

The dark matter halo is modelled with a generalised NFW model
(Navarro, Frenk, \& White 1996) with the central density profile slope
$\gamma$ left as a free parameter. This has a density profile given by
\be
\rho(r)=\frac{\rho_s}{(r/r_s)^{\gamma}(1+r/r_s)^{3-\gamma}}.
\ee
Here, $\rho_s$ is the halo normalisation and $r_s$ is a scale radius.
The volume mass density $\rho(r)$ must be projected to a 
surface mass density by integrating along the line of
sight. We generalise this surface mass density to an elliptical
distribution using the ellipse co-ordinate $\eta$. Following the
prescription given by Keeton (2002), the elliptical surface density is
used to calculate the deflection angle.

The parameters describing the lens model are then
the baryonic M/L $\Psi$, the position of the halo centre with respect
to the light centre $(x_h,y_h)$,  the halo
normalisation $\rho_s$, the central slope $\gamma$, the halo ellipticity
$e_h=1-b/a$ and the orientation of the semi-major axis to the
vertical in a counter-clockwise direction, $\theta_h$. We hold the
scale radius fixed at $r_s=40 \,h_{65}^{-1}$kpc $(\cong 9'' \,\, @ \,
z=0.485, \,\, \Omega=0.3, \,\,\Lambda=0.7)$ to match that expected from
simulations by Bullock et al. (2001) for a galaxy of equal mass and
redshift of the lens galaxy in 0047$-$2808. The fitted lens model
is insensitive to the choice of $r_s$.

Confidence contours are obtained on the $\Psi,\gamma$ plane by
minimising over the remaining 5 parameters at regular grid points in
the plane. We use a modified version of Powell's method (Press et
al. 2001) for the minimisation. The ring is masked with an annulus to
remove insignificant sky pixels (see Figure 2).

\section{Results}

Figure 2 shows the results of the minimisation. The confidence regions
on $\Psi$ and $\gamma$ are shown in the plot on the left. In order not
to bias the lens solution, this minimisation was not regularised
(ie. $\lambda$ was set to 0). We used a source plane size of $0.5''
\times 0.5''$ with an adaptive pixelisation scheme dependent on the
lens magnification (see Dye \& Warren, 2003). To $95\%$ confidence, we
measure an inner slope of $\gamma=0.87^{+0.69}_{-0.61}$ and a baryonic
M/L of $\Psi=3.05^{+0.45}_{-0.86} \,h_{65}\,M_{\odot}/L_{B\odot}$.
The dashed lines show the corresponding confidence levels obtained by
Koopmans \& Treu (2003) in analysing a combination of Keck
spectroscopy and the same HST observations as those here.  Their
constraints were the observed velocity dispersion profile and a
lensing estimate of the total mass enclosed by the Einstein ring.
Clearly, by using the full information content of the ring image, the
semi-linear method gives significantly better constraints.  Most
notably, we have been able to constrain the baryonic M/L without the
need for dynamical measurements. The lens galaxy is offset from the
local fundamental plane by a factor 0.37dex (Koopmans and Treu
2003). Correcting by this factor our derived M/L is in
remarkable agreement with the local average value for ellipticals of
$7.3\pm 2.1\,h_{65}\,M_{\odot}/L_{B\odot}$ (Gerhard et al. 2001).

The other minimised lens model parameters are
$(x_h,y_h)=(0.07'',0.03'')$, $e_h=0.180\pm0.02$ and $\theta_h=41.7\pm
2.7^{\circ}$, which give $\chi^2_{\nu}=0.942 \pm 0.040$.  The centre
of the halo is well aligned with the centre of the light whereas the
orientation is marginally inconsistent at the $\sim 2.5\sigma$
level. The halo is rounder than the light distribution with an
ellipticity of approximately half the value of $e=0.31$ for the Sersic
profile.  Finally, the halo is found to contribute 35\% of the total
{\em projected mass} within a circular region of radius of $1.16''$
traced by the ring.

\begin{figure}[h]
\plotone{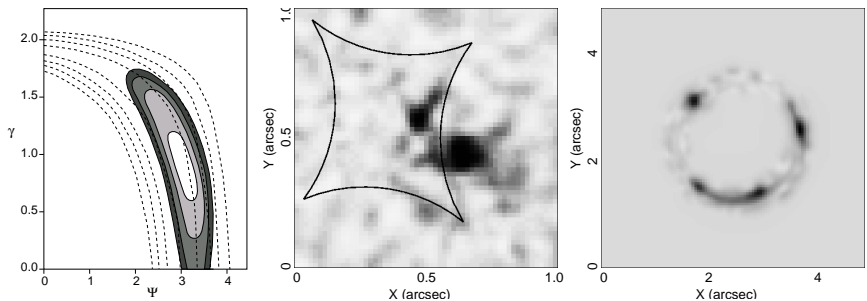}
\caption{\small {\em Left:} 68\%, 95\%, 99\% \& 99.9\% 
1$-$parameter confidence levels on inner slope $\gamma$ and baryonic M/L,
$\Psi$ (in units of $\,h_{65}\,M_{\odot}/L_{B\odot}$). Dashed lines
are same confidence levels from analysis by Koopmans \& Treu (2003)
using dynamics + lensing. {\em Middle:} Regularised reconstructed
source pair showing lens caustic (note different scale).  {\em Right:}
Image of regularised source in masked annular region.}
\end{figure}

The middle plot of Figure 2 shows the reconstructed source. To best
resolve the source morphology we performed a regularised inversion,
fixing the lens model parameters to those obtained above in the
unregularised solution. $\lambda$ was set to the value
Tr($\bsf{F})/$Tr($\bsf{R}$) to weight both matrices in equation (1)
equally. Although this gives a ring image (Figure 2, right) with a
$\chi^2$ worse than the unregularised case, the reconstructed source
allows identification of two distinct objects. This agrees with the
findings of Wayth et al. (2003) and illustrates the importance of
using a non-parametric source: Using a single analytic source would
have biased the minimised lens model in this case.

\section{Summary \& Discussion}

We have applied our semi-linear method to HST observations of the
Einstein ring system 0047$-$2808 to determine the lens galaxy mass
profile. Modelling the lens as a baryonic Sersic profile + point
source nested in a dark matter generalised NFW halo, we find a M/L of
$3.05^{+0.45}_{-0.86} \,h_{65}\,M_{\odot}/L_{B\odot}$ and an inner
slope $\gamma=0.87^{+0.69}_{-0.61}$ (95\% CL). This M/L was obtained
without any dynamical measurements. The reconstructed source surface
brightness distribution shows two distinct source objects,
highlighting the need for non-parametric sources to obtain unbiased
lens mass profiles.

One of the main predictions of the Cold Dark Matter (CDM) model is
that halo mass profiles are cuspy. Recent CDM simulations show that
the inner slope should lie somewhere in the range $1 \simless \gamma
\simless 1.5$ (Navarro, Frenk, \& White 1996; Moore et al. 1999). Our
measurement of $\gamma$ for 0047$-$2808 is not inconsistent with
these.  Nevertheless there are two important factors which must be
considered. The first is that the simulations are only just starting
to reliably probe scales of $\sim$ few kpc. The most recent
simulations seem to indicate that $\gamma$ continues to decrease at
smaller radii with no sign of reaching a convergent value (Power et
al. 2003; see also Navarro et al. in these proceedings). For complete
reassurance in comparing these predictions with observational
constraints, higher resolution simulations are needed.  The second
complication is that the majority of simulations model only pure dark
matter distributions.  The effect of a baryonic potential well alters
the shape of the dark matter halo in a non-trivial way. Clearly, these
shortcomings must be addressed before CDM can be tested with
confidence.


\begin{references}

\reference Bullock, J.S., Kollat, T.S., Sigad, Y., Somerville, R.S.,
Kravtsov, A.V., Klypin, A.A., Primack, J.R., Deckel, A., 2001, \mnras,
321, 598

\reference Dye, S. \& Warren S.J., 2003, ApJ, submitted

\reference Gerhard, O., Kronawitter, A., Saglia, R.P., \& Bender, R.,
2001, AJ, 121, 1936

\reference Hoekstra, H., Yee, H.K.C., Gladders, M.D., 2003,
astro-ph/0306515

\reference Keeton, C.R., 2002, astro-ph/0102341

\reference Kochanek, C.S., Keeton, C.R., \& McLeod, B.A., 2001, \apj,
547, 50

\reference Koopmans, L.V.E. \& Treu, T., 2003, ApJ, 583, 606

\reference Moore, B., Quinn, T., Governato, F., Stadel, J., Lake, G,
1999, MNRAS, 310, 1147

\reference Navarro, J.F., Frenk, C.S., \& White S.D.M., 1996, \apj,
462, 563

\reference Power, C., Navarro, J.F., Jenkins, A., Frenk, C.S., White,
S.D.M., Springel, V., Stadel, J., Quinn, T., 2003, \mnras, 338, 14

\reference Press, W.H., Teukolsky, S.A., Vetterling, W.T., Flannery,
B.P., 2001, 'Numerical Recipes in Fortran 77, 2nd Edition', Cambridge
University Press

\reference Sersic, J.L., 1968, Cordoba, Argentina: Observatorio
Astronomico (1968)

\reference Wallington, S., Kochanek, C.S., \& Narayan, R., 1996, \apj,
465, 64

\reference Warren, S.J., Hewett, P.C., Lewis, G.F., M{\o}ller, P.,
Iovino, A., Shaver, P.A., 1996, \mnras, 278, 139

\reference Warren, S.J. \& Dye, S., 2003, \apj, 590, 673

\reference Wayth, R.B., Warren, S.J., Lewis, G.F., Hewett, P.C., 2003,
\mnras, in prep

\end{references}
\end{document}